\documentclass[twocolumn]{aastex63}

\usepackage{amsmath}
\usepackage{algorithmic}

\received{2021 September 15}
\revised{2021 November 24}
\accepted{2021 December 9}

\newcommand{\kms}{  {km~s$^{-1}$}}
\newcommand{\dege}{$^{\circ}$}
\newcommand{\degee}{$^{\circ}$ }

\submitjournal{ApJL}

\shorttitle{Multipoint interplanetary coronal mass ejections}
\shortauthors{M\"ostl et al. }

\begin{document}

   

   


\title{Multipoint interplanetary coronal mass ejections observed with Solar Orbiter, BepiColombo, Parker Solar Probe, Wind and STEREO-A}

\correspondingauthor{Christian M\"ostl}
\email{christian.moestl@oeaw.ac.at}

\author[0000-0001-6868-4152]{Christian M\"ostl}
\affiliation{Space Research Institute, Austrian Academy of Sciences, Schmiedlstraße 6, 8042 Graz, Austria}
\affiliation{Institute of Geodesy, Graz University of Technology, Steyrergasse 30, 8010 Graz, Austria}

\author[0000-0002-6273-4320]{Andreas J. Weiss}
\affiliation{Space Research Institute, Austrian Academy of Sciences, Schmiedlstraße 6, 8042 Graz, Austria}
\affiliation{Institute of Geodesy, Graz University of Technology, Steyrergasse 30, 8010 Graz, Austria}
\affiliation{Institute of Physics, University of Graz, Universit\"atsplatz 5, 8010 Graz, Austria}

\author[0000-0002-6362-5054]{Martin A. Reiss}
\affiliation{Space Research Institute, Austrian Academy of Sciences, Schmiedlstraße 6, 8042 Graz, Austria}
\affiliation{Institute of Geodesy, Graz University of Technology, Steyrergasse 30, 8010 Graz, Austria}

\author[0000-0001-9024-6706]{Tanja Amerstorfer}
\affiliation{Space Research Institute, Austrian Academy of Sciences, Schmiedlstraße 6, 8042 Graz, Austria}

\author[0000-0003-2021-6557]{Rachel L. Bailey}
\affil{Zentralanstalt f\"ur Meteorologie und Geodynamik, Hohe Warte 38, 1190 Vienna, Austria}

\author[0000-0002-1222-8243]{J\"urgen Hinterreiter}
\affiliation{Space Research Institute, Austrian Academy of Sciences, Schmiedlstraße 6, 8042 Graz, Austria}
\affil{Institute of Physics, University of Graz, Universit\"atsplatz 5, 8010 Graz, Austria}

\author[0000-0002-2507-7616]{Maike Bauer}
\affiliation{Space Research Institute, Austrian Academy of Sciences, Schmiedlstraße 6, 8042 Graz, Austria}
\affil{Institute of Physics, University of Graz, Universit\"atsplatz 5, 8010 Graz, Austria}

\author[0000-0003-1137-8220]{David Barnes}
\affiliation{RAL Space, Rutherford Appleton Laboratory, Harwell Campus, Didcot, OX11 0QX, UK}

\author[0000-0001-9865-9281]{Jackie A. Davies}
\affiliation{RAL Space, Rutherford Appleton Laboratory, Harwell Campus, Didcot, OX11 0QX, UK}

\author[0000-0002-0843-8045]{Richard A. Harrison}
\affiliation{RAL Space, Rutherford Appleton Laboratory, Harwell Campus, Didcot, OX11 0QX, UK}

\author[0000-0002-1390-4776]{Johan L. Freiherr von Forstner}
\affiliation{Institute of Experimental and Applied Physics, University of Kiel, Kiel, Germany}
\affiliation{now at: Paradox Cat GmbH, Munich, Germany}

\author[0000-0001-9992-8471]{Emma E. Davies}
\affiliation{Space Science Center, Institute for the Study of Earth, Oceans, and Space, University of New Hampshire, Durham, NH, USA}

\author[0000-0001-7894-8246]{Daniel Heyner}
\affiliation{Technical University of Braunschweig, Braunschweig, Germany}

\author[0000-0002-7572-4690]{Tim Horbury}
\affiliation{Department of Physics, Imperial College London, London, UK}

\author[0000-0002-1989-3596]{Stuart D. Bale}
\affiliation{Physics Department and Space Sciences Laboratory, University of California, Berkeley, CA, USA}

\begin{abstract}
We report the result of the first search for multipoint in situ and imaging observations of interplanetary coronal mass ejections (ICMEs) starting with the first Solar Orbiter (SolO) data in 2020 April - 2021 April. A data exploration analysis is performed including visualizations of the magnetic field and plasma observations made by the five spacecraft SolO, BepiColombo, Parker Solar Probe (PSP), Wind and STEREO-A, in connection with coronagraph and heliospheric imaging observations from STEREO-A/SECCHI and SOHO/LASCO. We identify ICME events that could be unambiguously followed with the STEREO-A heliospheric imagers during their interplanetary propagation to their impact at the aforementioned spacecraft, and look for events where the same ICME is seen in situ by widely separated spacecraft. We highlight two events: (1) a small streamer blowout CME on 2020 June 23 observed with a triple lineup by PSP, BepiColombo and Wind, guided by imaging with STEREO-A, and (2) the first fast CME of solar cycle 25 ($ \approx 1600$ \kms) on 2020 November 29 observed in situ by PSP and STEREO-A. These results are useful for modeling the magnetic structure of ICMEs and the interplanetary evolution and global shape of their flux ropes and shocks, and for studying the propagation of solar energetic particles. The combined data from these missions are already turning out to be a treasure trove for space weather research and are expected to become even more valuable with an increasing number of ICME events expected during the rise and maximum of solar cycle 25.
\end{abstract}

\keywords{Unified Astronomy Thesaurus concepts: Solar coronal mass ejections (310); Solar storm (1526); Solar wind (1534); Solar physics (1476); Interplanetary physics (827); Interplanetary magnetic fields (824); Interplanetary shocks (829); Heliosphere (711); Space weather (2037); Solar-planetary interactions (1472); Solar system (1528). \vspace{2 cm}}

\section{Introduction} \label{sec:intro}

Two major problems in space weather research concern (1) establishing a connection between solar eruptions observed in the solar corona and their signatures measured in situ by spacecraft in interplanetary space, and (2) finding events where more than one spacecraft experiences the eruptive event at different locations in the inner heliosphere. Magnetic field observations by magnetometers are currently the only available method to measure the magnetic field structure of the flux ropes in interplanetary coronal mass ejections (ICMEs), but vastly undersample the enormous size of ICMEs, giving us only an incomplete picture. Thus, the global shape and structure of ICMEs is still poorly understood, despite four decades of research since the inception of this field of study with \cite{Burlaga_1981}. This poor understanding directly affects our ability to forecast the arrival and geomagnetic effects of CMEs \citep[e.g.][]{kay_2018,vourlidas_2019,luhmann_2020_review}.

With a fleet of spacecraft in operation since 2020 April, formed by the four deep-space probes Solar Orbiter (SolO), Parker Solar Probe (PSP), BepiColombo, and STEREO-A, in conjunction with one of the L1 spacecraft such as Wind, we are now in a new, golden era of multipoint solar wind observations. Right after the beginning of Solar Orbiter (SolO) magnetometer observations, a stealth CME was observed in situ by SolO at 0.8~AU, together with Wind at 1~AU and BepiColombo near Earth \citep{Davies_2021, weiss_2021_solo, forstner_2021, okane_2021}. These and case studies of other events \citep[e.g.][]{winslow2021} already gave a glimpse into the possibilities for studying ICMEs in the upcoming years with not only in situ lineups, but also adding completely novel viewpoints with SolO going to high latitudes after 2025 \citep{mueller_2020}. The unprecedented close orbits of Parker Solar Probe lead to possible in situ double crossings of CMEs \citep{moestl_2020}, additionally guided by imaging \citep{liewer_2019}.

This paper presents the first systematic search for multipoint in situ and imaging events since the start of SolO, in combination with the other four aforementioned missions. We focus on two events to highlight the possibilities for deriving new knowledge about ICMEs using these types of events. To address problem (1), the connection between imaging and in situ CME events, we employ the STEREO-Ahead Heliospheric Imagers \citep[HI,][]{eyles2009}, essentially following our previous studies \citep{Moestl_2014,Moestl2017}. The STEREO-HI observations allow us to seamlessly connect solar and interplanetary observations of CMEs with the Self-Similar-Expansion geometrical fitting model with 30\dege{} half width \citep[SSEF30,][]{davies2012}. The results from SSEF30 are available for a large cohort of the CMEs observed with STEREO-HI so far \citep[$\approx 1600 $ events,][]{barnes2019}. More sophisticated approaches exist such as ELEvoHI \citep{amerstorfer2018, hinterreiter_2021} or 3D density shell models \citep[e.g.][]{wood2017}, but they have been applied to a much smaller set of events so far.

In order to make progress on problem (2), finding multipoint in situ events in collective heliospheric observations, we connect ICMEs observed at two or more spacecraft that are separated by up to $\approx 60$\degee heliospheric longitude \citep{bothmer1998}, often aided by HI. Given the low rate of CMEs during solar minimum and in the early rising phase~\citep[e.g.][]{webb2012_review}, these connections are relatively straightforward as there is often only a single candidate CME present to explain the in situ signatures at the different spacecraft. If a CME is directed between the longitude of STEREO-A and a direction of about 120\dege{} west of STEREO-A \citep{lugaz2012}, the CME can be tracked with HI out to 1~AU, which often leads to an unambiguous connection of different in situ observations.

To better understand the CME magnetic structure and interplanetary evolution, various scales of in situ spacecraft separations are needed in all three dimensions \citep[radial, longitudinal, latitudinal;][]{lugaz2018}. Previous catalogs of multipoint ICME events have been shown by \cite{Good_2019} and \cite{Salman_2020} for STEREO and Wind in conjunction with Venus Express (2006--2014) and MESSENGER (2004--2015). Until SolO starts to travel to higher latitudes starting in 2025, all spacecraft are confined to a few degrees around the ecliptic and solar equatorial planes, so until then all separations are significant only in the longitudinal and radial directions.

Here, we present a catalog of multipoint ICME observations to advance our understanding of the evolution of CME flux ropes and solar energetic particles \citep[e.g.][]{mitchell2021}. The catalog is publicly available online at \url{https://www.helioforecast.space/lineups}.

\section{Methods} \label{sec:methods}

\begin{figure*}[ht!]
\noindent\includegraphics[width=\linewidth]{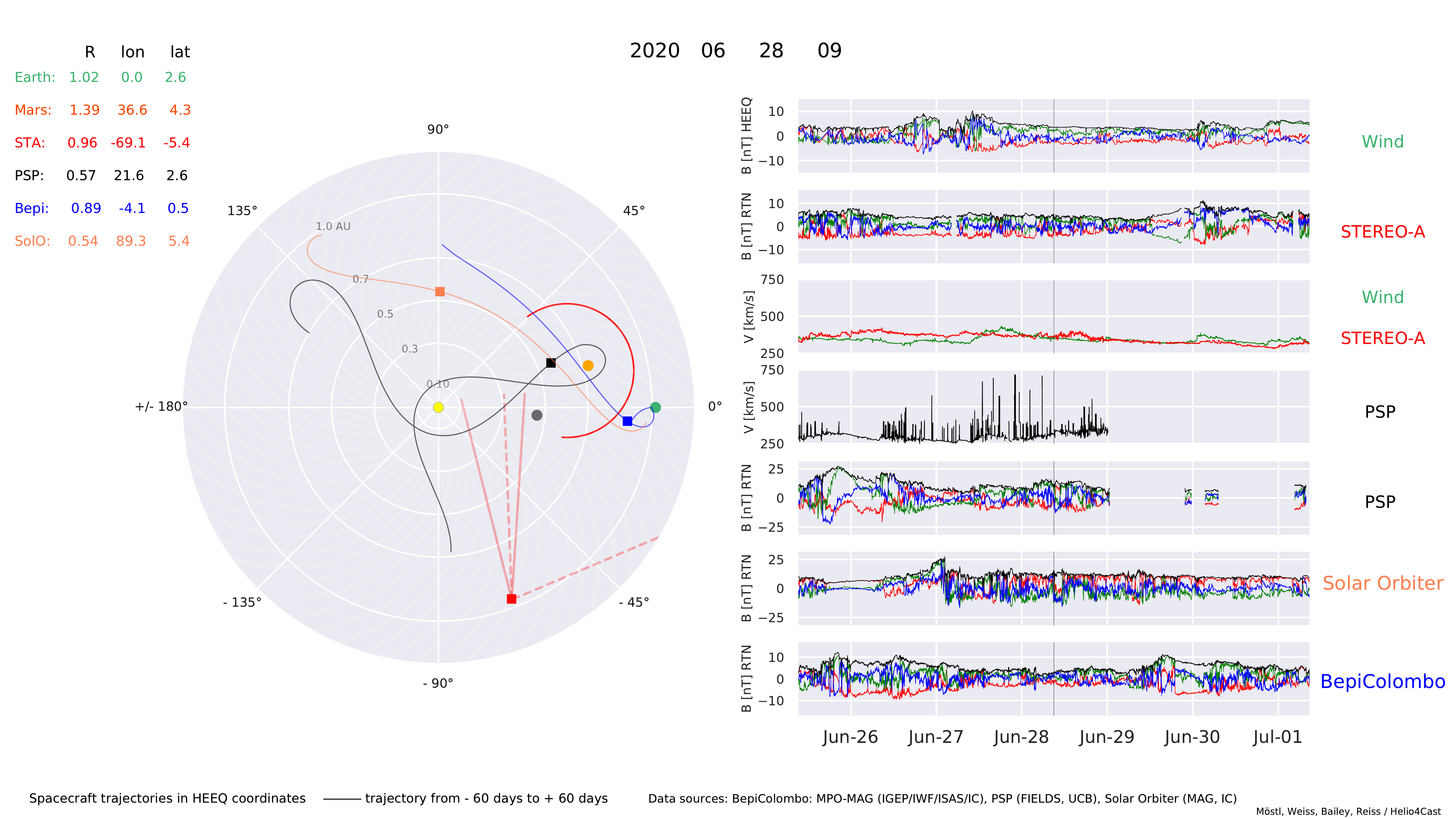}
\caption{Frame of an animation that is used to identify multipoint ICME events and their connection to the Sun with CME kinematics derived from heliospheric imaging. The 2020 June triple in situ CME event is shown. Left panel: the spacecraft positions with PSP (black square), BepiColombo (blue square), STEREO-Ahead (red square) and Earth (green circle). The field of views by the STEREO-A HI instruments are given as solid (HI1) and dashed (HI2). The red semi-circle indicates a CME front resulting from SSEF30 modeling. Right panel: from top to bottom, with labels on the right side of the plots: Wind magnetic field data in HEEQ coordinates ($B_t$ black, $B_x$ red, $B_y$ green, $B_z$ blue), STEREO-A magnetic field in RTN ($B_t$ black, $B_r$ red, $B_t$ green, $B_n$ blue), bulk plasma speed at STEREO-A and Wind, PSP bulk plasma speed, and then the PSP, SolO and BepiColombo magnetic field observations, in RTN coordinates. \newline
The animation begins on 2020 April 10 and runs through 2021 April, 14 in one-hour intervals. The real-time duration of the animation is 9 minutes and 52 s.  (An animation of this figure is available at \url{https://youtu.be/cfDzFtvo3nQ}.)}
\label{fig:anim_frame}
\end{figure*}

We have manually identified multipoint CME events observed by imaging and in situ instruments from 2020 April 1 to 2021 April 1. To this end, we generated movies for this time range from daily or monthly quicklook data that are available from SOHO/LASCO C2/C3 \citep{brueckner1995} and the STEREO-A SECCHI COR2/HI1/HI2 coronagraphs and heliospheric imagers \citep{howard2008a}. We additionally created an animation that includes (1) the heliospheric position of the five spacecraft, Wind (at the Sun--Earth L1 point), SolO, PSP, BepiColombo, and STEREO-Ahead (STA), (2) CME fronts modeled after observations by STEREO-A HI \citep{eyles2009}, and (3) in situ magnetic field and plasma observations. The animation is also available on YouTube (\url{https://youtu.be/cfDzFtvo3nQ}). All movies and the animation are included in the data repository for this paper on figshare (see the Acknowledgements).

\textbf{Figure~\ref{fig:anim_frame}} shows a frame of this animation for event search and selection. On the left side, the spacecraft positions are indicated, their future and past trajectory for $-60$ to $+60$~ days, as well as the field of view of the STEREO-A heliospheric imager cameras HI1 (solid) and HI2 (dashed). On the right side, in situ magnetic field observations taken by these instruments are shown: Wind MFI \citep{lepping1995}, STA IMPACT \citep{Luhmann_2008}, PSP FIELDS \citep{bale2016_fields},  SolO MAG \citep{horbury_2020}, and BepiColombo MPO/MAG \citep{heyner2021}. Plasma data is available from Wind SWE \citep{ogilvie1995}, STA PLASTIC \citep{galvin2008} and PSP SWEAP \citep{kasper2016_sweap}. The magnetic field data coverage is essentially continuous at Wind and STEREO-A, and has some gaps at the other three spacecraft, which, however, does not strongly impact our search for multipoint ICMEs.   

Our search is aided by the following catalogs: (1) The HELCATS HIGeoCAT \citep{barnes2019}, which provides directions and speeds of CMEs observed with the STEREO-A heliospheric imagers (HIA). These are obtained by tracking the CME in Jmaps near the ecliptic and subsequently applying geometrical modeling with the SSEF30 technique \citep{davies2012}. From this we generate kinematics that let us visualize the propagation and impacts of CMEs in Fig.~\ref{fig:anim_frame}. (2) The CME arrival times in the events in HIGeoCat at various planets and spacecraft are given in the HELIO4CAST ARRCAT arrival catalog. (3) The HELIO4CAST ICMECATv2.0 catalog \citep{moestl_2020} is a living catalog that is updated with ICME observations from the five spacecraft, and includes over 800 events so far since 2007. The selection criteria for the ICMECAT are similar to \cite{nieves_2018_wind}, including clear identifications of ICMEs that show magnetic obstacles which range from cleanly rotating flux ropes to complex ejecta. Furthermore, the LASCO/CDAW CME catalog \citep{yashiro2004} is referenced for parameters of the CMEs close to the Sun, but we limit its usage here to the two events we investigate more thoroughly.

In general, events are added to the multipoint ICME list that have a clear HI to in situ connection, which means that the modeled arrival is $\Delta_t<0\pm 24$~hr within the in situ arrival \citep{Moestl2017}. However, this criterion is not strict, and CME events that closely miss (within $< 20$\dege{)} the in situ spacecraft and that arrive approximately within $\Delta_t$ are also included. Given the solar minimum to early rising phase conditions, only two events are mergers, the rest are isolated CMEs.

\section{Results} \label{sec:results}


\subsection{Lineup event table} \label{sec:lineup_table}

\begin{deluxetable*}{ccccccccc}\label{tab:insitu_hi}
\tabletypesize{\footnotesize } 
\tablecaption{\textbf{Multipoint imaging and in situ coronal mass ejection events}}
\tablewidth{2.0\linewidth}
\tablehead{
\colhead{Event} & \colhead{Start time [UT]} & \colhead{Spacecraft} & \colhead{ R [AU]} & \colhead{S/c long. [$^{\circ}]$} & \colhead{CME long. [$^{\circ}$]} & \colhead{Speed [km~s$^{-1}$]}
& \colhead{$B_t$ max [nT]} & \colhead{$B_z$ min [nT]}} 
\startdata
     1 & 2020-04-15 20:49 &           HIA &   0.967 &            -75.4 &               -6 &          339 &          \ldots&          \ldots\\
     1 & 2020-04-19 05:06 &  SolarOrbiter &   0.808 &             -4.0 &               \ldots&           \ldots&        21.2 &       -19.2 \\
     1 & 2020-04-20 01:34 &          Wind &   0.996 &              0.2 &               \ldots&          346 &        16.2 &       -15.1 \\
     1 & 2020-04-20 03:09 &   BepiColombo &   1.011 &             -1.3 &               \ldots&           \ldots&        16.1 &       -14.4 \\\hline
     2 & 2020-05-26 01:29 &           HIA &   0.966 &            -72.3 &               36 &          497 &          \ldots&          \ldots\\
     2 & 2020-05-28 12:45 &  SolarOrbiter &   0.564 &             28.2 &               \ldots&           \ldots&        28.2 &       -26.6 \\
     2 & 2020-05-29 21:20 &          Wind &   1.005 &             -0.2 &               \ldots&          350 &        14.5 &       -12.4 \\
     2 & 2020-05-29 15:27 &   BepiColombo &   0.977 &             -5.2 &               \ldots&           \ldots&        13.1 &       -11.5 \\\hline
     3 & 2020-05-27 04:49 &           HIA &   0.966 &            -72.2 &              -47 &          315 &          \ldots&          \ldots\\
     3 & 2020-06-03 07:54 &      STEREO-A &   0.965 &            -71.5 &               \ldots&          373 &         9.9 &        -4.6 \\\hline
     4 & 2020-06-23 00:49 &           HIA &   0.963 &            -69.6 &                 16 &      290 &      \ldots &        \ldots \\
     4 & 2020-06-25 11:39 &           PSP &   0.523 &             19.9 &            \ldots &         302 &    27.4 &    -23.1 \\
     4 & 2020-06-29 09:44 &   BepiColombo &   0.881 &             -3.9 &            \ldots &     \ldots &    11.4 &     -7.8 \\
     4 & 2020-06-30 01:12 &          Wind &   1.007 &             -0.2 &            \ldots &       332 &     6.4 &     -3.3 \\\hline
     5 & 2020-07-06 06:09 &           HIA &   0.962 &            -68.4 &              -61 &          407 &          \ldots&          \ldots\\
     5 & 2020-07-09 14:17 &      STEREO-A &   0.962 &            -68.0 &               \ldots&          374 &        15.3 &       -13.9 \\\hline
     6 & 2020-07-09 08:49 &           HIA &   0.962 &            -68.1 &               47 &          493 &          \ldots&          \ldots\\
     6 & 2020-07-13 20:18 &          Wind &   1.006 &             -0.2 &               \ldots&          377 &         9.4 &        -8.9 \\\hline
     7 & 2020-08-16 14:09 &           HIA &   0.958 &            -64.8 &              -47 &          467 &          \ldots&          \ldots\\
     7 & 2020-08-19 21:11 &      STEREO-A &   0.958 &            -64.6 &               \ldots&           \ldots&        19.6 &       -13.4 \\\hline
     8 & 2020-09-08 16:49 &           HIA &   0.957 &            -63.1 &               41 &          237 &          \ldots&          \ldots\\
     8 & 2020-09-12 10:23 &           PSP &   0.473 &             13.0 &               \ldots&          350 &        19.4 &        -9.6 \\\hline
     9 & 2020-09-30 18:49 &           HIA &   0.956 &            -61.5 &              -14 &          322 &          \ldots&          \ldots\\
     9 & 2020-10-05 06:52 &          Wind &   0.991 &              0.2 &               \ldots&          353 &        10.7 &        -8.6 \\\hline
    10 & 2020-10-06 15:29 &           HIA &   0.956 &            -61.1 &               65 &          619 &          \ldots&          \ldots\\
    10 & 2020-10-08 10:27 &   BepiColombo &   0.694 &             79.5 &               \ldots&           \ldots&        15.2 &         -11 \\
    10 & 2020-10-10 04:07 &   BepiColombo &   0.698 &             80.4 &               \ldots&           \ldots&        12.3 &        -4.5 \\\hline
    11 & 2020-10-26 19:29 &           HIA &   0.957 &            -59.7 &              -33 &          384 &          \ldots&          \ldots\\
    11 & 2020-11-01 11:54 &          Wind &   0.984 &              0.1 &               \ldots&          352 &         9.4 &        -7.4 \\\hline
    12 & 2020-11-29 13:30 &           GCS &      \ldots&              \ldots &             -106 &   1637       &          \ldots&          \ldots\\
    12 & 2020-12-01 02:22 &           PSP &   0.804 &            -96.8 &               \ldots&           \ldots&          39 &        14.6 \\
    12 & 2020-12-01 07:28 &      STEREO-A &   0.959 &            -57.6 &               \ldots&           \ldots&        15.9 &       -12.1 \\\hline
    13 & 2020-11-29 16:09 &           HIA &   0.959 &            -57.7 &              -23 &          537 &          \ldots&          \ldots\\
    13 & 2020-12-01 10:49 &           HIA &   0.959 &            -57.6 &              -37 &          652 &          \ldots&          \ldots\\
    13 & 2020-12-03 18:04 &      STEREO-A &   0.960 &            -57.5 &               \ldots&           \ldots&        10.1 &        -9.3 \\\hline
    14 & 2021-02-11 01:29 &           HIA &   0.966 &            -55.8 &              -24 &          380 &          \ldots&          \ldots\\
    14 & 2021-02-15 18:58 &          Wind &   0.977 &              0.1 &               \ldots&          354 &        11.2 &         -10 \\\hline
    15 & 2021-02-20 16:48 &           HIA &   0.967 &            -55.7 &              -26 &          423 &          \ldots&          \ldots\\
    15 & 2021-02-23 10:34 &      STEREO-A &   0.967 &            -55.7 &               \ldots&           \ldots&        11.4 &       -10.8 \\
    15 & 2021-02-24 04:08 &          Wind &   0.979 &              0.2 &               \ldots&          481 &         9.4 &        -6.8 \\\hline
    16 & 2021-03-22 08:48 &           HIA &   0.967 &            -55.0 &               -8 &          353 &          \ldots&          \ldots\\
    16 & 2021-03-26 18:00 &          Wind &   0.989 &              0.2 &               \ldots&          373 &         6.9 &        -4.7 \\\hline
    17 & 2021-03-31 00:08 &           HIA &   0.967 &            -54.7 &              -28 &          397 &          \ldots&          \ldots\\
    17 & 2021-04-04 21:52 &          Wind &   0.992 &              0.2 &               \ldots&          320 &           9 &        -8.4 \\
    17 & 2021-04-06 01:27 &      STEREO-A &   0.967 &            -54.4 &               \ldots&           \ldots&        12.5 &       -10.2 \\\hline
\enddata
\tablecomments{Column `Start time' is either the first appearance in the STEREO-A heliospheric imager (HIA) or the in situ arrival time of the ICME. This is equal to the shock arrival time if a shock is present. For event \#12, it is the time of first appearance in SOHO/LASCO/C3. Column `Spacecraft' is the in situ spacecraft where the CME arrives. When this is set to HIA, the parameters given by the location and derived from observations by HIA are shown in that row. For event \#12, the parameters from GCS fitting are quoted. `R' is the heliocentric distance in AU and `S/c long.' the spacecraft longitude in HEEQ coordinates at the given start time. `CME long.' is the CME longitude in HEEQ derived from SSEF30 modeling based on CME tracking with HIA. The `Speed' is the in situ observed mean bulk speed of the ICME over the full ICME interval (sheath and magnetic obstacle), or for HIA the speed given by SSEF30 modeling. The last two columns are the in situ observed maximum total magnetic field $B_t$ and the minimum $B_z$ component in the magnetic obstacle (either in HEEQ coordinates for Wind or RTN coordinates for all other spacecraft).}
\end{deluxetable*}

\textbf{Table \ref{tab:insitu_hi}} is the main product of this study. We have found 17 events in total that are observed by more than one spacecraft. In fact, for 16 out of 17 events in the list, the connection between HIA and at least one in situ spacecraft could be made. The single event in the list that is not connected to the solar wind closer to the Sun with HIA, and observed at just two in situ observatories, is the fast CME first appearing in SOHO/LASCO/C3 on 2020 Nov 29 13:30 UT (event \#12), which was propagating mostly east of STA. It could be clearly observed in situ at both PSP, 39\degee east of STA, and by STA itself on 2020 Dec 1. There are 2 merger events, for which the connection between heliospheric imaging and in situ measurements is not straightforward. This is why we have stated 2 candidate CMEs in HIA for an impact at STA (event \#13) and 2 ICMEs at BepiColombo for which there was only one clear candidate CME in HI (event \#10). For these events, further investigations concerning CME-CME interaction may be in order.

As seen in the speed column, most CMEs had slow speeds around 300 to 500\kms, in situ as well as derived from HI modeling, and maximum magnetic field strengths in the 10--15 nT range at 1~AU, both typical for solar minimum. The event with in situ measurements at the closest distance to the Sun is event \#8 observed by PSP at 0.47~AU (for Solar Orbiter event \#2, at 0.57~AU). Note that PSP had already observed a CME at 0.25 AU, but this was not a multipoint event \citep[e.g.][]{nieves_chinchilla_2020, weiss_2021_fit}. These events are particularly suited for investigations of the radial CME evolution.

The cleanest multipoint flux rope event so far is event \#1, a stealth CME on 2020 April 15--20, already well studied \citep[e.g.][]{Davies_2021,forstner_2021, weiss_2021_solo, okane_2021}. We now briefly touch on two particularly outstanding events in our first multipoint CME list that we think merit deeper investigations in future studies. These sections are intended to showcase the significance of ICME multipoint events for the research on solar eruptions.

\subsection{2020 June 23 CME} \label{sec:june}


\begin{figure*}[ht!]
\noindent\includegraphics[width=\linewidth]{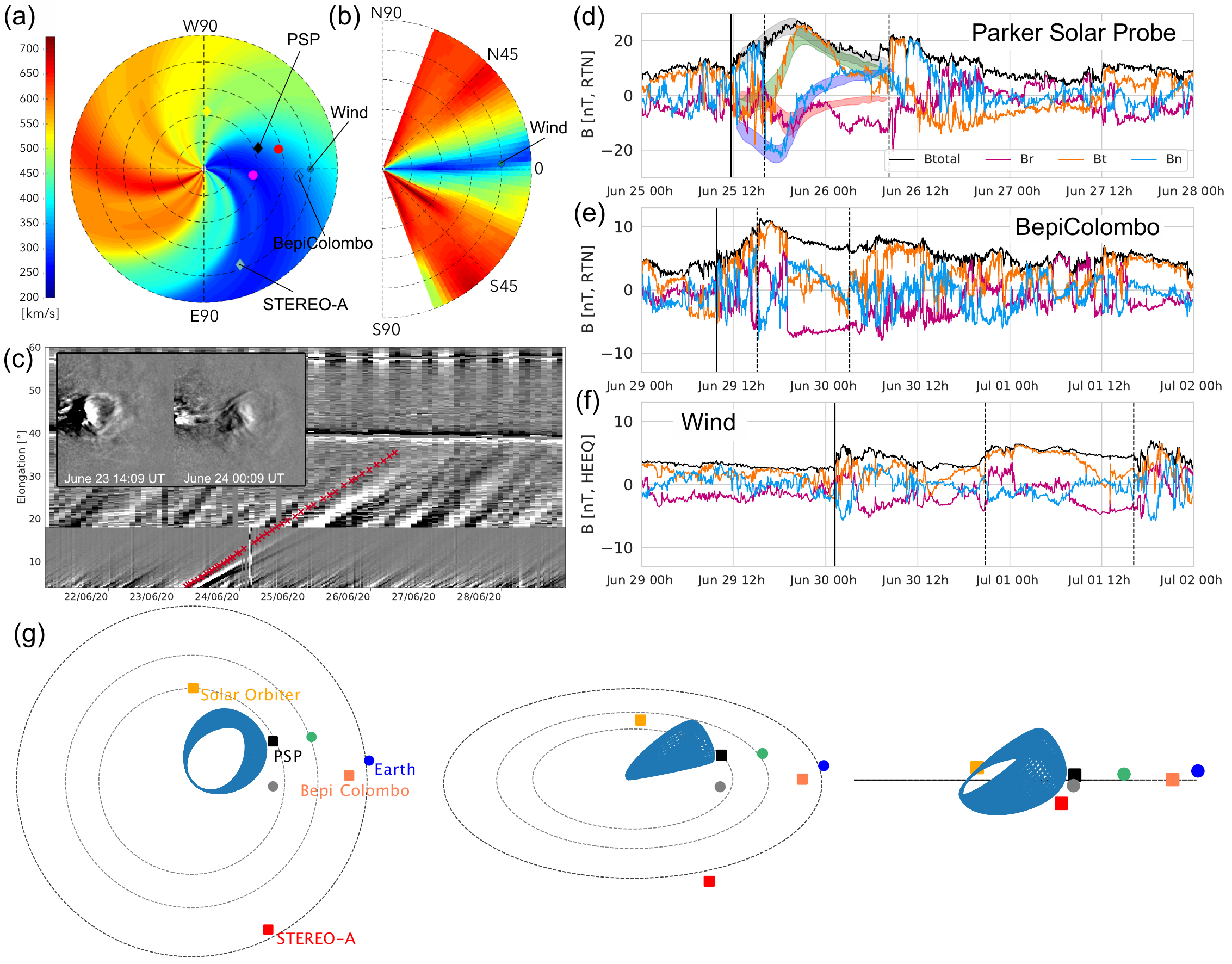}
\caption{CME event on 2020 June 23. (a) WSA/THUX background solar wind in the solar equatorial plane and (b) in a meridional plane at the Earth longitude for 2020 June 27 01:00 UT. (c) STEREO-A HI1/2 Jmap with HI1 images of the CME, and the HELCATS time-elongation track of the CME front overlaid as red crosses. (d) In situ magnetic field data at PSP, and (e) BepiColombo in RTN coordinates and (f) Wind in HEEQ  coordinates (total field magnitude in black, $B_r$ magenta, $B_t$ orange, $B_n$ blue). (g) Visualization of 3DCORE flux rope modeling for the 2020 June 25 ICME at PSP in three different viewing angles, with the location of the magnetic flux rope and positions of the spacecraft given for the time 2020 June 25 07:00 UT. In panels (d-f) solid vertical lines indicate ICME start times, and dashed lines the boundaries of the magnetic obstacle. In panel (d) the results from 3DCORE ABC--SMC fitting method are given as shaded areas for each magnetic field component. Note the time shift of 4 days in panels (e) and (f) compared to panel (d). }
\label{fig:june_event}
\end{figure*}

\textbf{Figure~\ref{fig:june_event}} is an overview of event \#4, a triple point in situ and heliospheric imaging event of a very slow streamer blowout CME. 
\textbf{Figs.~\ref{fig:june_event}a,b} show the ambient solar wind speed from the Wang--Sheeley--Arge /  Tunable Heliospheric Upwind Extrapolation model \citep[WSA/THUX,][]{reiss_2020_2}, \textbf{Fig.~\ref{fig:june_event}c} the Jmap from HIA, with an overlay of two sample images from HIA, and the in situ magnetic field data at the three spacecraft in \textbf{Figs.~\ref{fig:june_event}d,e,f}.

This CME is very faint in LASCO, and cataloged as a poor event with an initial speed of only 80\kms. But it is clearly observed by STA COR2 from about 70\dege{} east of the Sun-Earth line to propagate straight in the ecliptic, with a very slow rise starting early on 2020 June 22. It entered HI1A on 2020 June 23 00:49 UT with an edge-on flux rope morphology, indicating a low inclination magnetic flux rope orientation. The application of SSEF30 modeling shows that it was directed 16\degee (HEEQ) west of the Sun-Earth-line, propagating with a speed of 290\kms.


PSP was situated at 0.52 AU and at 19.9\degee longitude, only 4\degee west of the CME apex direction of 16\degee given by SSEF30, at the in situ observed ICME arrival time of June 25 11:39 UT. The predicted arrival time of this CME at PSP is June 25 22:30 UT, about 11 hours later than observed in situ, still well within the average error bars of $\pm 17$ hours for this type of modeling \citep{Moestl2017}. Given that the CME is isolated with no other candidate CME erupting in the days before or after this event, the connection between in situ and imaging data is clearly established for PSP. \textbf{Figure~\ref{fig:june_event}d} shows the in situ magnetic field data \citep[see also][]{palmerio2021_pred}. There is no shock but an increase in the total magnetic field before a magnetic obstacle is seen between June 25 15:59 UT and June 26 08:15 UT. In its first part, this is a clean, right-handed and low-inclination South-West-North (SWN, in RTN $B_n < 0$, $B_t >0$, $B_n > 0$) magnetic flux rope (MFR) with a maximum total field of 27.4 nT. It exhibits a tail-like structure after around June 26 02:00 UT where the field components remain essentially constant. This is often interpreted as a sign of flux rope erosion by magnetic reconnection with the ambient wind \citep[e.g.][]{dasso2007, pal2021}.

\textbf{Figure~\ref{fig:june_event}g} shows a visualization of the 3DCORE results for a single-spacecraft fit to the PSP in situ magnetic field data. We fitted the 3DCORE flux rope model \citep[for details, see][]{weiss_2021_fit, weiss_2021_solo} to the MFR interval with an Approximate Bayesian Computation - Sequential Monte Carlo (ABC-SMC) algorithm, which results in a direction of $38 \pm 7$\dege{} longitude (22\dege{} west of the SSEF30 direction) and $7 \pm 8$\dege{} latitude (HEEQ), an inclination of $34 \pm 11$\dege{} to the solar equatorial plane and a cross-section aspect ratio of $3.1 \pm 0.9$. Prior to the fit, we set the launch time to June 23 02:54 UT at 15 solar radii (based on STEREO-A COR2 images), and restricted the aspect ratio to $< 5$. The results confirm the HIA images showing a low or only moderately inclined flux rope orientation.

The BepiColombo magnetic field data (\textbf{Figure~\ref{fig:june_event}e}) show the ICME arrival on June 29 09:44 UT, when the spacecraft was positioned at 0.88 AU and -3.9\dege{} east of Earth. Here, the discrepancy between the arrival time predicted from SSEF30 (June 28 20:25 UT) and the in situ arrival time is about 13 hours, but now the CME arrives later in situ than given by the SSEF30 model (opposite to PSP). The magnetic signature at Bepi can also be interpreted as SWN but it is more distorted and the radial component $B_r < 0$ is more pronounced. It could be that BepiColombo, 24\dege{} in longitude east of PSP, sees the flank of this ICME. The tail part at the back of the magnetic obstacle is much more pronounced, with a clear discontinuity separating the rotating field in the beginning with the latter part exhibiting almost constant field components.


Arriving at Wind at 1.01 AU on June 30 01:12 UT (\textbf{Figure~\ref{fig:june_event}f}), a magnetic obstacle is seen with a field showing $B_n < 0$ and $B_t >0$, which is reminiscent of the southwest part of the flux ropes seen earlier by BepiColombo and PSP but missing the northward field part. The arrival time given by SSEF30 is June 29 07:31 UT, 18 hours earlier as observed in situ.


Several interesting conclusions can be drawn from these multi-spacecraft observations: 

\begin{itemize}

\item The magnetic field data can be interpreted as radial observations of a progressive erosion of the flux rope up to 1 AU, making the magnetic flux rope lose flux and thereby altering the geoeffective $B_n$ component.

\item The SSEF30 modeling assumes a circular front shape and constant speed. In this event, the constant speed assumption seems not to be violated: the in situ arrival speed at Wind of the ICME is 326\kms, consistent with the speed of 302\kms{} at PSP and the 290\kms{} derived from HIA. This means that the discrepancies, ranging from $-11$ to $+18$ hr between predicted and observed arrival times, must be attributed to a strong deformation of the CME front from a circular shape \citep[see][]{hinterreiter_2021}.

\item The slow rise of the CME,  the slow interplanetary propagation speed of the CME and the ambient wind speed on the order of only 300\kms, as well as a flank impact at Earth leads to an extremely long travel time of this CME of 7 days from entering HIA to the arrival at 1 AU at Wind. There is no other CME following in the next days that can explain these in situ observations.

\end{itemize}

\subsection{2020 November 29 CME} \label{sec:nov}


\begin{figure*}[ht!]
\noindent\includegraphics[width=\linewidth]{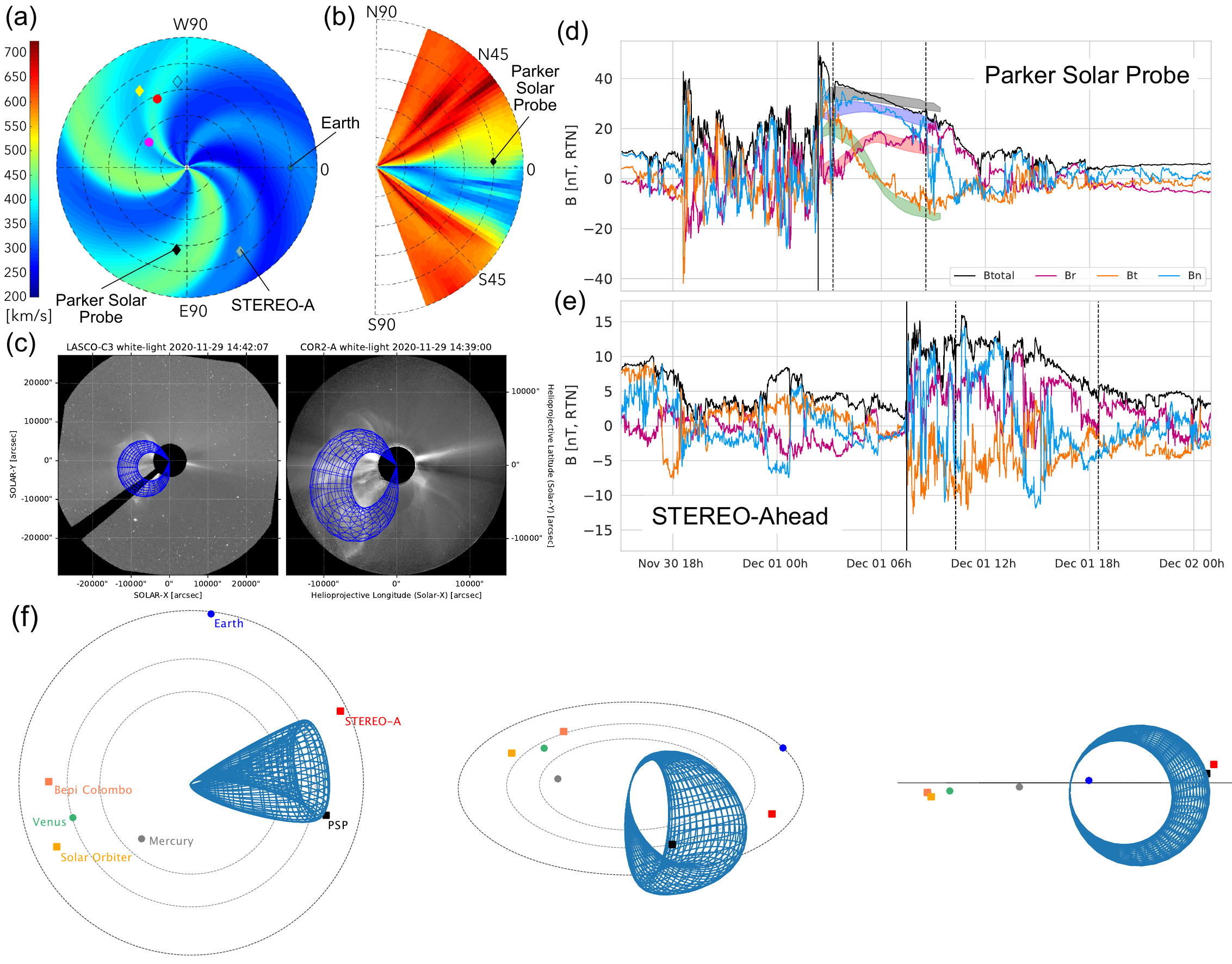}
\caption{2020 Nov 29 CME. (a) WSA/THUX background solar wind in the solar equatorial plane and (b) in a meridional plane at the PSP longitude on 2020 Dec 1 12:00 UT. (c) GCS fit to STEREO-A COR2 and SOHO LASCO/C3, (d) Parker Solar Probe and (e) STEREO-A in situ magnetic field data. (f) Visualization of 3DCORE flux rope modeling for the 2020 Dec 1 ICME at Parker Solar Probe in three different viewing angles, with the location of the magnetic flux rope and positions of the spacecraft given for the time 2020 Dec 1 02:00 UT. In panel (d) the results from 3DCORE ABC-SMC fitting to the PSP flux rope are given as shaded areas for each magnetic field component.}
\label{fig:event_dec}
\end{figure*}

\textbf{Figure~\ref{fig:event_dec}} shows event \#12, a double in situ event at STA and PSP. This is the first fast CME observed in solar cycle 25. It entered SOHO/LASCO C3 on 2020 November 29 13:30 UT and propagated through the C3 field of view with a projected linear speed of 2077\kms{} (from the LASCO CME catalog). It was also observed by STA/COR2 heading to mainly east, and was even partly observed by STA/HI1, although the CME was directed eastward and outside the field of view of STA/HI1, and so we omit the SSEF30 results here as they are not representative of the global CME parameters.

\textbf{Figure~\ref{fig:event_dec}a} gives the background wind speed in the solar equatorial plane as given by WSA/THUX \citep{reiss_2020_2} at the arrival time of the ICME shock at PSP (2020 December 1 02:22 UT). It is seen that the CME propagates into slightly elevated solar wind speed (450\kms{}). However, the meridional cut in \textbf{Figure~\ref{fig:event_dec}b} at the longitude of PSP indicates that this high speed stream was confined to northern latitudes, and so we expect that this high inclination CME should be strongly deformed during interplanetary propagation as it encounters solar wind with different speeds at different latitudes \citep[e.g.][]{owens2006}.




To obtain the CME parameters as it propagates through the coronagraphs, we performed Graduated Cylindrical Shell \citep[GCS,][]{thernisien_2006} modeling with STEREO-A COR2 and SOHO/LASCO C3 images (\textbf{Figure~\ref{fig:event_dec}c}). We obtain a CME direction of $-106 \pm 10$\degee longitude (HEEQ), close to quadrature to Earth, and $-17 \pm$ 5\dege{} latitude (HEEQ) \citep[for a discussion of error bars see][]{forstner_2021}. The half angle is 60\dege{}, consistent with it being a very wide and fast CME. A linear fit to the GCS measurements for the apex between 6.27 and 14.87 solar radii (from 13:39 to 14:39 UT) gives a radial speed of 1637~km~s$^{-1}$, about 400 km~s$^{-1}$ slower than the speed given in the LASCO catalog. The CME intensity is very structured, so it is difficult to distinguish face-on versus edge-on view features \citep{thernisien_2006}, but the high angular width seen in quadrature strongly suggests a high inclination flux rope. We find a good fit only to be produced by a high tilt or inclination, for which we set the highest possible value of 90\degee to the solar equatorial plane.

\textbf{Figure~\ref{fig:event_dec}d} shows the magnetic field data taken by PSP FIELDS. PSP at 0.804 AU is situated at a longitude of -97\degee (HEEQ), which is just 9\degee west from the GCS CME initial direction. The field data perfectly confirm the high inclination flux rope. We see a clear rotation of the magnetic field from west ($B_t> 0$) to north ($B_n > 0$) to east ($B_t > 0$), making this a right handed and high-inclination WNE flux rope \citep{mulligan1998}. \textbf{Figure~\ref{fig:event_dec}f} gives a 3D view of the 3DCORE fit results. Here we obtain a direction of $-76 \pm 10$\dege{} longitude (30\dege{} west of the GCS direction) and $-10 \pm 12$\dege{} latitude (HEEQ), an inclination of $80 \pm 6$\dege{} to the solar equatorial plane and a cross-section aspect ratio of $3.0 \pm 0.8$.

\textbf{Figure~\ref{fig:event_dec}e} shows the magnetic field at STEREO-A, which was situated close to the Sun-Earth L5 point at -58\dege, at 0.959 AU, 48\degee west of the CME apex given by GCS and 39\degee west of PSP. Clearly, the high-inclination flux rope does not extend over this range of longitude, as no flux rope signatures are seen and only a shock followed by a sheath region and a complex magnetic obstacle is observed.  This event is thus a very good example for the influence on the background wind on the possible deformation and evolution of a fast CME and for the longitudinal extension of a high inclination flux rope.

\section{Conclusions}

We have presented the results of the first multipoint and imaging CME event search in the Solar Orbiter, BepiColombo and Parker Solar Probe era. There are essentially three classes of events available. (1) CME events with the connection between imaging and in situ observations at more than one in situ spacecraft are clearly the most valuable for deciphering the global CME magnetic structure. Here, we have direction and morphology information through imaging and are able to see the otherwise undiagnosed magnetic field in situ along the spacecraft trajectories, which is crucial for understanding the ICME flux rope configuration \citep[e.g.][]{Burlaga_1981,lugaz2018,Davies_2021,weiss_2021_solo, forstner_2021}. (2) Events that have coverage with heliospheric imaging and that encounter a single in situ spacecraft are also valuable, as we need as many events as possible to link the CME morphology in images to the flux rope inclination, density, and speed of the in situ ICME observation. (3) Multipoint in situ events without heliospheric imaging, but available coronagraph imaging, can also be used for multipoint fitting and testing CME propagation models to better understand the CME and shock structure and shape. 

We have done a short analysis of two outstanding events from the list, the CME on 2020 June 23 falling into category (1) and the CME on 2020 November 29 into category (3). The 2020 June 23 event is an opportunity to study ICME erosion happening along the flank of a very slow streamer blowout CME. The 2020 November 29 event is an example of a fast CME, rarely observed in recent years, that impacts Parker Solar Probe centrally and just grazes STEREO-A. Both events show a consistency between the morphology in coronagraph and HI images and their internal magnetic field structure, which is seen as a possibility to derive their southward pointing magnetic field intervals days in advance for space weather forecasting purposes.

Many more such favorable configurations are expected to arise in solar cycle 25. Multipoint ICME events provide key observations for our improved understanding of the global CME magnetic structure and shape, the origin of CME magnetic fields, their interplanetary evolution, the interactions with other CMEs and the ambient solar wind, the evolution of shock waves, the propagation of solar energetic particles, and for real time $B_z$ prediction.

Our presented catalog is intended as a living catalog to inform the research community of novel multipoint ICME observations. These are exciting opportunities that should enable a leap in our understanding of solar eruptions for spaceweather research and forecasting.

\acknowledgments
We have benefited from the availability of Solar Orbiter, STEREO, Wind, Parker Solar Probe and BepiColombo data, and thus would like to thank the instrument teams and data archives for their data distribution efforts. C.M., A.J.W., R.L.B, M.A., T.A., M.B. and J.H. thank the Austrian Science Fund (FWF): P31265-N27, P31521-N27, P31659-N27. E.D. gratefully acknowledges the support of NASA grant 80NSSC19K0914 and partial support of NASA STEREO grant 80NSSC20K0431. J. v. F. thanks the German Space Agency (Deutsches Zentrum für Luft- und Raumfahrt e.V., DLR) for their support of his work on the Solar Orbiter EPD team under grant 50OT2002. This research used version 1.1.0 (doi: 10.5281/zenodo.3604353) of the SunPy open source software package \citep{mumford_2020_sunpy}. This research made use of Astropy (\url{http://www.astropy.org}) a community-developed core Python package for Astronomy \citep{astropy_2018}. The code for GCS modeling in python is available at \url{https://github.com/johan12345/gcs_python} (doi: 10.5281/zenodo.5084818). STEREO-HI data processing has been done with \url{https://github.com/maikebauer/STEREO-HI-Data-Processing}. The LASCO CME catalog is available at \url{https://cdaw.gsfc.nasa.gov/CME_list}, the HELIO4CAST ICMECATv2.0 at  \url{https://www.helioforecast.space/icmecat}, the HELIO4CAST ARRCAT at \url{https://www.helioforecast.space/arrcat}, and the HELCATS HIGeoCAT at \url{https://www.helcats-fp7.eu}.
 The code to produce the multipoint ICME list in Table 1 is available as a jupyter notebook at \url{https://github.com/helioforecast/Papers/tree/master/Moestl2021_multipoint} and the catalog, figures, quicklook movies and animations for this paper are archived at \url{https://doi.org/10.6084/m9.figshare.15134745}. The multipoint ICME catalog can be accessed at \url{https://helioforecast.space/lineups}. We thank the reviewer for helpful suggestions that improve the manuscript.




\bibliography{chrisbib}

\end{document}